\documentclass[12pt,letterpaper]{article}
\usepackage{amsmath}
\usepackage{amssymb}
\usepackage{verbatim}
\usepackage{physics}
\usepackage{qcircuit}

\newcommand{\R}{\mathbb{R}}
\newcommand{\N}{\mathbb{N}}
\newcommand{\Z}{\mathbb{Z}}
\newcommand{\C}{\mathbb{C}}
\newcommand{\spn}{\operatorname{span}}

\usepackage[backend=bibtex8]{biblatex}
\usepackage{braket}
\bibliography{bibliography.bib}

\title{Approximating Unitary Preparations of Orthogonal Black Box States}
\author{Joshua Alan Cook}

\begin{document}
	
\maketitle

\begin{abstract}
	In this paper, I take a step toward answering the following question: for m different small circuits that compute m orthogonal n qubit states, is there a small circuit that will map m computational basis states to these m states without any input leaving any auxiliary bits changed. While this may seem simple, the constraint that auxiliary bits always be returned to 0 on any input (even ones besides the m we care about) led me to use sophisticated techniques. I give an approximation of such a unitary in the m = 2 case that has size polynomial in the approximation error, and the number of qubits n.
\end{abstract}

\section{Introduction}

The problem I focus on in this paper is: given orthogonal states \(\ket{\psi}\) and \(\ket{\phi}\), and black box unitary $U$ and $V$ for preparing them from \(\ket{0^n}\), construct a unitary $W$ that takes \(\ket{0^n}\) to \(\ket{\psi}\) and \(\ket{10^{n-1}}\) to \(\ket{\phi}\). In particular, it might use auxiliary bits initialized to 0, but these bits will always be returned to 0 for any input. We show several special cases where we can easily do this exactly, and show an algorithm to approximate $W$ in time polynomial in the estimation error. To do this, we also prove some related results such as unitary preparations of rotations between orthogonal states.

This problem is in contrast to more straightforward problems in that we want this to return the auxiliary bits to 0 for EVERY input, not just $\ket{0^n}$ and $\ket{10^{n-1}}$ The straightforward algorithm simply uses one auxiliary bit to see which one of the two input states we are in, and cleans up the control bit based off the output. Unfortunately, while this algorithm works on the two inputs we care about, it doesn't work on all of them, in particular \(\ket{\phi}\), and \(\ket{\psi}\) are still problems. While this can still be used in some special cases for \(\ket{\phi}\) and \(\ket{\psi}\), it doesn't work in general.

If the input and output states are orthogonal, then the problem is easy. Similarly, if the input and output states span the exact same sub space, this is also exactly doable. But in the general case, we only find that we can construct an $\epsilon$ approximation of a unitary that gives results these results in time polynomial in \(\frac{1}{\epsilon}\). Exact computation and computation with more than 2 output states is still open.

I will allow access to the following 4 black box operations with our given unitaries: application, adjoint application, controlled application, and controlled adjoint application. The intuitive reason for allowing these applications is that in most physical implementations of our black box unitary (a small quantum circuit in particular), these operations would all be available.

In solving this problem, I also develop a couple other tools that may be of interest. Given an operation that computes \(\ket{\psi}\) from \(\ket{0^n}\) that cleans up all auxiliary bits on that input, I give an operation that approximately computes the component of \(\ket{\psi}\) orthogonal to \(\ket{0^n}\) cleaning up all auxiliary bits on any input. Further, given such an operation as above, I give a procedure to approximate an operation that computes \(\ket{\psi}\) and does not effect any inputs orthogonal to \(\ket{\psi}\) and \(\ket{0^n}\).

\subsection{Motivation}

Lots of quantum operations are known to exist in theory, but in practice actually making these quantum operations is difficult even if parts of the problem are classically easy. One such problem is Q-Sampling, where we may know how to prepare an appropriate probability distribution over the states, but only as a mixed state, not as a pure quantum state. In general, we often have the issue that when computing a desired result, we use auxiliary bits and leave them in a modified state. This can leave our result in a mixed state that won't have the desired interference for a larger algorithm.

I want to provide tools for preparing quantum states without disturbing any auxiliary bits. This way, the result of a pure input state is always still a pure state.

I examine the problem of just merging different operations for computing orthogonal output states and creating one operation that computes one on one orthogonal input, and the other on another. This is clearly doable in theory, but straightforward algorithms will cause auxiliary bits to not be reset on some inputs. If this is a subroutine in a larger algorithm that may feed in one of these inputs to our algorithm, the result will be a mixed state which may prevent future interference that was expected.

In practice, no practical application of these results are known. But these results do give us a class of unitary operations that should exist and gives a way to efficiently compute them.

\subsection{Related Work}

Previous work has investigated how to run quantum algorithms with auxiliary qubits in an arbitrary quantum state without disturbing them \cite{Chi2005}. In particular showing Simon and Shor's algorithm can be done using any auxiliary bits in any arbitrary quantum state without disturbing them. But that paper focused on minimizing the need for special working space bits, while this paper focuses on making sure our algorithms leave no garbage on any input.

The gentle measurement lemma is a huge component of the main result. Gentle measurement lemmas have been introduced in various forms across quantum computing literature \cite{And1999}, \cite{Mark2015}, \cite{Mark2011}. Most of these focus on the disturbance introduced by measurement operators if the the result of the measurement is close to certain. In particular, if one can predict the outcome of measurement \(\Lambda\) with probability \(1 - \epsilon\), then the trace distance to the resulting state after the measurement is \(O(\sqrt{\epsilon})\). We need essentially this same result except that we don't actually have a measurement.

Thus we prove a related result. Basically, we need a result that says if some computation $U$ prepares some bits that are used as inputs in some mixed state to another quantum algorithm acting on another register where most of the results yield the other register to approximately compute the same result, then applying \(U^{-1}\) will mostly clear the input register.

A recent paper greatly expands on what we know about gentle measurements on quantum states by relating them to differential privacy \cite{Scot2019}.

\subsection{Notation}

First, I need to solidify notation a little. For n bit, orthogonal quantum states \(\ket{\psi}\) and \(\ket{\phi}\), we have black box access to unitary $U$ such that \(U\ket{0^n} = \ket{\psi}\) and unitary $V$ such that \(V \ket{0^n} = \ket{\phi}\).

First, I note that we can controlled apply a gate based on if a state is the all 0s state. This can be done using the standard techniques of using n bits to calculate the and not of all the bits, CNOTing on the result, then decomputing. For notation, I call the gate that computes this control \(M_{\ket{0^n}}\) so that

$$
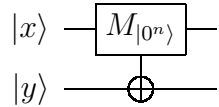
\Qcircuit @C=1em @R=.7em {
	\lstick{\ket{x}} & \gate{M_{\ket{0^n}}} \qwx[1] & \qw \\
	\lstick{\ket{y}} & \targ & \qw \\
}$$

Computes state \(\ket{x}\ket{y \oplus (x == 0^n)}\). Similarly, for any unitary U preparing state \(\ket{\psi}\), we can define \(M_{\ket{\psi}}\) as:

$$
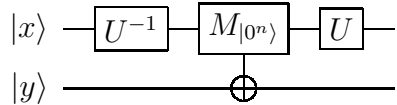
\Qcircuit @C=1em @R=.7em {
	\lstick{\ket{x}} & \gate{U^{-1}} & \gate{M_{\ket{0^n}}} \qwx[1] & \gate{U} & \qw \\
	\lstick{\ket{y}} & \qw & \targ & \qw & \qw \\
}$$

which will just apply the control if \(\ket{x} = \ket{\psi}\), but will do nothing to any orthogonal state.

\section{Swapping Orthogonal States}

Define \(W = V U^{-1}\). This takes \(\ket{\psi}\) and gives \(\ket{\phi}\). Then we just control apply $W$ conditioned on the \(\ket{\psi}\) state and \(W^{-1}\) conditioned on the \(\ket{\phi}\) state. This will require auxiliary bits to do but we know which input went to which, so we can clean up the axilary bits. Note we need \(\ket{\psi} \perp \ket{\phi}\) because otherwise applying the controlled not to prepare the auxilary bit would change the state of the other.

I actually give a slightly simpler circuit here though that is easier to prove works. This one takes \(\ket{\psi}\) to \(\ket{\phi}\), and in the process of cleaning its auxiliary bit turns the bit indicating that we should take \(\ket{\phi}\) to \(\ket{\psi}\). See:

$$
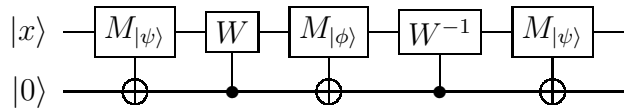
\Qcircuit @C=1em @R=.7em {
	\lstick{\ket{x}} & \gate{M_{\ket{\psi}}} \qwx[1] & \gate{W} & \gate{M_{\ket{\phi}}} \qwx[1] & \gate{W^{-1}} & \gate{M_{\ket{\psi}}} \qwx[1] & \qw \\
	\lstick{\ket{0}} & \targ & \ctrl{-1} & \targ & \ctrl{-1} & \targ & \qw \\
}$$

To see that this works, see that for any input state \(\ket{x}\), there exists state \(\ket{y}\) orthogonal to states \(\ket{\psi}\) and \(\ket{\phi}\) and scalars \(\alpha, \beta, \gamma\) such that:

\[\ket{x} = \alpha \ket{\psi} + \beta \ket{\phi} + \gamma \ket{y}\]

Then our circuit performs the following:

\[
\begin{split}
& (\alpha \ket{\psi} + \beta \ket{\phi} + \gamma \ket{y})\ket{0} \\
\rightarrow & \alpha \ket{\psi}\ket{1} + \beta \ket{\phi}\ket{0} + \gamma \ket{y}\ket{0} \\
\rightarrow & \alpha W\ket{\psi}\ket{1} + \beta \ket{\phi}\ket{0} + \gamma \ket{y}\ket{0} \\
= & \alpha \ket{\phi}\ket{1} + \beta \ket{\phi}\ket{0} + \gamma \ket{y}\ket{0} \\
\rightarrow & \alpha \ket{\phi}\ket{0} + \beta \ket{\phi}\ket{1} + \gamma \ket{y}\ket{0} \\
\rightarrow & \alpha \ket{\phi}\ket{0} + \beta W^{-1}\ket{\phi}\ket{1} + \gamma \ket{y}\ket{0} \\
= & \alpha \ket{\phi}\ket{0} + \beta \ket{\psi}\ket{1} + \gamma \ket{y}\ket{0} \\
\rightarrow & \alpha \ket{\phi}\ket{0} + \beta \ket{\psi}\ket{0} + \gamma \ket{y}\ket{0} \\
= & (\alpha \ket{\phi} + \beta \ket{\psi} + \gamma \ket{y})\ket{0} \\
\end{split}
\]

This swaps \(\ket{\psi}\) and \(\ket{\phi}\) as desired.

Note under the hood, this is applying two of our black box unitaries on each step, which is 5 steps in total, which is still a constant number. This means that with only constant overhead, we can always assume that the black box unitaries that prepare an orthogonal pure state only swap the two states and do nothing else.

\subsection{Hardening Clean Computation of Orthogonal State To A Unitary}

Now I give an algorithm that takes a unitary $U$ on \(n+m\) bits such that \(U\ket{0^n}\ket{0^m} = \ket{\psi}\ket{0^m}\) where \(\bra{\psi}\ket{0^n} = 0\), and constructs a unitary operation transformation $U'$ on n bits which swaps \(\ket{0^n}\) with \(\ket{\psi}\) and leaves all orthogonal states unchanged. See that this is different then the last case because the input $U$ might produce garbage on any input other than \(\ket{0^n}\). Nevertheless, the algorithm is actually the same.

We can use the same procedure used above, because \(U^{-1}\) also leaves no garbage on the state \(\ket{\psi}\), and we can actually measure whether we are in the state \(\ket{\psi}\) without disturbing any other state. This can be done by just looking at the measurement circuit:

$$
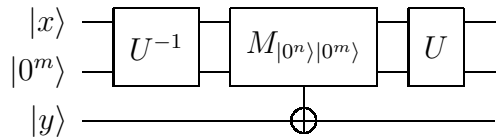
\Qcircuit @C=1em @R=.7em {
	\lstick{\ket{x}} & \multigate{1}{U^{-1}} & \multigate{1}{M_{\ket{0^n}\ket{0^m}}}  & \multigate{1}{U} & \qw \\
	\lstick{\ket{0^m}} & \ghost{U^{-1}} & \ghost{M_{\ket{0^n}\ket{0^m}}} \qwx[1] & \ghost{U} & \qw \\
	\lstick{\ket{y}} & \qw & \targ & \qw & \qw \\
}$$

This can easily be seen to take \(\ket{x}\ket{0^m}\ket{y}\) to

\[\ket{x}\ket{0^m}\ket{y \oplus (x == \psi)}\]

by writing \(\ket{x}\) as

\[\ket{x} = \alpha\ket{0^n} + \beta \ket{\psi} + \gamma \ket{z}\]

Then the computation gives

\[\begin{split}
& (\alpha\ket{0^n} + \beta \ket{\psi} + \gamma \ket{z})\ket{0^m}\ket{y} \\
\rightarrow & U^{-1}(\alpha\ket{0^n} + \beta \ket{\psi} + \gamma \ket{z})\ket{0^m}\ket{y}\\
= &  U^{-1}(\alpha\ket{0^n} + \gamma \ket{z})\ket{0^m}\ket{y} + \beta \ket{0^{n}}\ket{0^m}\ket{y}\\
\rightarrow &  U^{-1}(\alpha\ket{0^n} + \gamma \ket{z})\ket{0^m}\ket{y} + \beta \ket{0^{n}}\ket{0^m}\ket{y \oplus 1}\\
\rightarrow &  U U^{-1}(\alpha\ket{0^n} + \gamma \ket{z})\ket{0^m}\ket{y} + \beta U \ket{0^{n}}\ket{0^m}\ket{y \oplus 1}\\
= &  (\alpha\ket{0^n} + \gamma \ket{z})\ket{0^m}\ket{y} + \beta \ket{\psi}\ket{0^m}\ket{y \oplus 1}\\
\end{split}\]

Then the swapping states algorithm above still works because it only ever applies $U$ to \(\ket{0^n}\) and \(U^{-1}\) to \(\ket{\psi}\) in addition to measurements of whether the state is \(\ket{0^n}\) or \(\ket{\psi}\), which I have shown none of which produce garbage.

This shows that it suffices to get an operation that cleanly computes state \(\ket{\psi}\) from any perpendicular state we can efficiently compute to get a unitary that swaps the two states, changes no other inputs, and produces no garbage.

\section{Preparation of Phases Between Orthogonal States}

Now, for arbitrary \(\alpha\) and \(\beta\) with \(\|\alpha\|^2 + \|\beta\|^2 = 1\), we want a unitary that takes \(\ket{0^n}\) and gives \(\alpha\ket{\psi} + \beta\ket{\phi}\).

To do this, we start with the observation that \(\ket{0^n} \perp U^{-1}\ket{\phi}\), since \(\bra{0^n} U^{-1} \ket{\phi} = \bra{\psi}\ket{\phi}\). So letting \(\ket{\omega} = U^{-1}\ket{\phi}\), let us instead prepare \(\alpha\ket{0^n} + \beta\ket{\omega}\), then we can apply U to this to get the result. Let \(W = U^{-1} V\) so that \(W \ket{0^n} = \ket{\omega}\).

Now the straight forward way to try this is to first prepare a control bit in the desired super position, controlled on it apply $U$, then based on the state now being \(\ket{\omega}\), flip the control bit back. This works on input \(\ket{0^n}\), but on input \(\ket{\omega}\), we actually get the auxiliary bit set to 1.

But this can easily be fixed. All we need to do is push the information about whether we are in state \(\ket{0^n}\) or \(\ket{\omega}\) into the control bit before we apply the rotation. Then we will still apply the rotation and clean up the input bit. We also need to take care to clean up the rotation, even on other orthogonal inputs. So, we just prepare a control qubit to only do this algorithm on the two important state, \(\ket{0^n}\) and \(\ket{\omega}\).

So let $R$ be our rotation such that \(R\ket{0} = \alpha \ket{0} + \beta \ket{1}\). Such a rotation can be made with arbitrary precision using standard techniques. Then our quantum circuit will just be:

$$
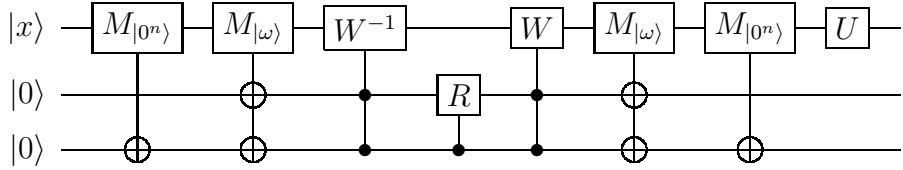
\Qcircuit @C=1em @R=.7em {
	\lstick{\ket{x}} & \gate{M_{\ket{0^n}}} \qwx[2] & \gate{M_{\ket{\omega}}} \qwx[2] & \gate{W^{-1}} & \qw & \gate{W} & \gate{M_{\ket{\omega}}} \qwx[2] & \gate{M_{\ket{0^n}}} \qwx[2] & \gate{U} & \qw \\
	\lstick{\ket{0}} & \qw & \targ & \ctrl{-1} & \gate{R} & \ctrl{-1} & \targ & \qw & \qw & \qw \\
	\lstick{\ket{0}} & \targ & \targ & \ctrl{-1} & \ctrl{-1} & \ctrl{-1} & \targ & \targ & \qw & \qw \\
}$$

Then similar to last time, for any \(\ket{x}\) and appropriately chosen scalars and orthogonal \(\ket{y}\):

\[\ket{x} = \gamma\ket{0^n} + \delta\ket{\omega} + \eta \ket{y} \]

Applying the circuit gives

\[
\begin{split}
& (\gamma\ket{0^n} + \delta\ket{\omega} + \eta \ket{y}) \ket{0}\ket{0} \\
\rightarrow & \gamma\ket{0^n}\ket{0}\ket{1} + \delta\ket{\omega}\ket{0}\ket{0} + \eta \ket{y}\ket{0}\ket{0}  \\
\rightarrow & \gamma\ket{0^n}\ket{0}\ket{1} + \delta\ket{\omega}\ket{1}\ket{1} + \eta \ket{y}\ket{0}\ket{0}  \\
= & (\gamma\ket{0^n}\ket{0} + \delta\ket{\omega}\ket{1})\ket{1} + \eta \ket{y}\ket{0}\ket{0}  \\
\rightarrow & (\gamma\ket{0^n}\ket{0} + \delta\ket{0^n}\ket{1})\ket{1} + \eta \ket{y}\ket{0}\ket{0}  \\
\rightarrow & (\gamma\ket{0^n}(\alpha\ket{0} + \beta\ket{1}) + \delta\ket{0^n}(\beta^*\ket{0} - \alpha^*\ket{1}))\ket{1} + \eta \ket{y}\ket{0}\ket{0}  \\
\rightarrow & (\gamma(\alpha\ket{0^n}\ket{0} + \beta\ket{\omega}\ket{1}) + \delta(\beta^*\ket{0^n}\ket{0} - \alpha^*\ket{\omega}\ket{1}))\ket{1} + \eta \ket{y}\ket{0}\ket{0}  \\
\rightarrow & \gamma(\alpha\ket{0^n}\ket{0}\ket{1} + \beta\ket{\omega}\ket{0}\ket{0}) + \delta(\beta^*\ket{0^n}\ket{0}\ket{1} - \alpha^*\ket{\omega}\ket{0}\ket{0}) + \eta \ket{y}\ket{0}\ket{0}  \\
\rightarrow & \gamma(\alpha\ket{0^n}\ket{0}\ket{0} + \beta\ket{\omega}\ket{0}\ket{0}) + \delta(\beta^*\ket{0^n}\ket{0}\ket{0} - \alpha^*\ket{\omega}\ket{0}\ket{0}) + \eta \ket{y}\ket{0}\ket{0}  \\
= & (\gamma(\alpha\ket{0^n} + \beta\ket{\omega}) + \delta(\beta^*\ket{0^n} - \alpha^*\ket{\omega}) + \eta \ket{y})\ket{0}\ket{0}  \\
\rightarrow & (\gamma(\alpha\ket{\psi} + \beta\ket{\phi}) + \delta(\beta^*\ket{\psi} - \alpha^*\ket{\phi}) + \eta U \ket{y})\ket{0}\ket{0}  \\
\end{split}
\]

Which when \(\gamma = 1\) gives the expected state \(\alpha\ket{\psi} + \beta\ket{\phi}\) and in all cases clears up the auxiliary bit to 0.

\section{Exact Preparation of Special Cases}

Now we get to the big question. We want to construct a unitary that takes \(\ket{0^n}\) to \(\ket{\psi}\) and \(\ket{10^{n-1}}\) to \(\ket{\phi}\).  Since the input and output states are orthogonal, there should be some transformation that does this. But can we do it with just the transformations we have been given? Yes, and in a couple special cases we can do it exactly.

\subsection{All are Orthogonal}

If the input and output states are all orthogonal to one another, then by using the state swapping procedure above, we can construct the unitary. We can also do it the input and output span the same subspace.

\subsection{\(\spn(\ket{0^n}, \ket{10^{n-1}}) = \spn(\ket{\psi}, \ket{\phi})\)}

In this case we can do the obvious thing, just measure if we are in \(\ket{0^n}\) or \( \ket{10^{n-1}}\), and if we are, apply the appropriate transformation. Then clean up the auxiliary bits. Let \(V'\) be the helper that flips the first bit then applies $V$ so that \(V'\ket{10^{n-1}} = \ket{\phi} \).

The quantum circuit for this transformation would be

$$
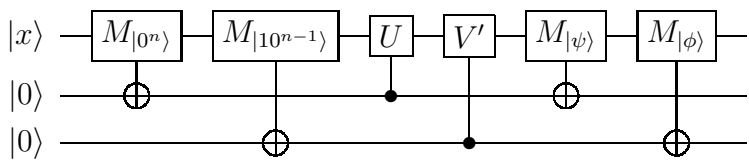
\Qcircuit @C=1em @R=.7em {
	\lstick{\ket{x}} & \gate{M_{\ket{0^n}}} \qwx[1] & \gate{M_{\ket{10^{n-1}}}} \qwx[2] & \gate{U} & \gate{V'} & \gate{M_{\ket{\psi}}} \qwx[1]  & \gate{M_{\ket{\phi}}} \qwx[2] & \qw \\
	\lstick{\ket{0}} & \targ & \qw & \ctrl{-1} & \qw & \targ & \qw & \qw \\
	\lstick{\ket{0}} & \qw & \targ & \qw & \ctrl{-2} & \qw & \targ & \qw \\
}$$

We can make a similar argument as before for the correctness of this algorithm, with for every \(\ket{x}\) there is appropriate orthogonal \(\ket{y}\) so that \(\ket{x} = \alpha \ket{0^n} + \beta \ket{10^{n-1}} + \gamma \ket{y}\). I leave this as an exercise, but note that this only works because \(\ket{y}\) is orthogonal to both \(\ket{0^n}, \ket{10^{n-1}}\) and \(\ket{\psi}, \ket{\phi}\), since \(\ket{\psi}, \ket{\phi}\) define the same subspace. If this were not true, then when we do the second measurement, some of \(\ket{y}\) would be changed and cause garbage in our auxiliary bits.

\subsection{Combining the Two}

Now we first realize that we can reduce this problem to the case where we want to swap \(\ket{10^{n-1}}\) with some special \(\ket{\omega}\) where \(\ket{0^n} \perp \ket{\omega}\) without effecting \(\ket{0^n}\). This is because \(U^{-1}\ket{\phi} = \ket{\omega} \perp \ket{0^n}\), so if we swap \(\ket{10^{n-1}}\) with \(\ket{\omega}\), we can apply $U$ to get the final, desired result.
	
Well, if \(\ket{10^{n-1}} = \ket{\omega}\), then actually we don't have to do anything, and if \(\ket{10^{n-1}} \perp \ket{\omega}\), we can do the swapping procedure described above. If we know that one of these two is true, then we can apply one of them. The challenge is determining which case this is. Well actually these states are very easy to distinguish: just run \(U^{-1} V\) on a random \(\ket{0^n}\) auxiliary bits and see if we get \(\ket{10^{n-1}}\). Then controlled on one of these, we apply one algorithm or the other, then uncompute the control bit.

But we still haven't covered the general case where \(|\bra{\omega}\ket{10^{n-1}}| \notin \{0, 1\}\): that is they differ by some other angle. But, we can handle the general case if we are okay with losing a little exactness by examining the properties of $U$ directly.

\section{General Case}

As noted earlier, we can reduce the general case to just the problem of taking state \(\ket{0^n}\) to some state \(\ket{\psi}\) and leaving all orthogonal states alone, given some unitary to take \(\ket{0^n}\) to \(\ket{\psi}\), which may modify orthogonal states. So we will instead solve this simpler problem given unitary $U$ to do the computation, knowing we can easily translate this solution back.

The general algorithm leverages the algorithm we developed earlier for making unitary rotations and requires two more tools:

\begin{enumerate}
	\item Estimate the angle of the rotation. We do this by running $U$ many times and summing up the number of the outcomes that were \(\ket{0^n}\). After many times, with high probability this will get us a decent sample of the rotation. After we are done using \(\theta\), leveraging the gentle measurement lemma, we can mostly clean the calculation of \(\theta\), resulting in a state with short trace distance from the desired state.
	
	\item Getting a unitary to compute the component of \(\ket{\psi}\), call it \(\ket{\psi'}\), that is orthogonal to \(\ket{0^n}\). This is the part that requires the most care.
	
	To do this, we apply $U$, then we will check whether we are in the \(\ket{0^n}\) state and apply the $U$ again, repeatedly until we have a very low amplitude on the \(\ket{0^n}\) state. Then, we use the known angle to clear up the working space bits we used when checking if the state was \(\ket{0^n}\) for all but the small amplitude we still ended up in the state \(\ket{0^n}\) after many conditional applications.
	
	We then use earlier algorithms to harden this procedure to be unitary, and then the rotation procedure to calculate \(\ket{\psi}\).
\end{enumerate}

Of course, the big difference between this case and the special cases described above is that we will not be computing the results exactly, and the number of queries we need to make to $U$ is a function of our error bound \(\epsilon\). The actual algorithm has 5 steps:

\begin{enumerate}
	\item Estimate the angle between \(\ket{0^n}\).
	
	\item Find the operation for taking \(\ket{0^n}\) to \(\ket{\omega'}\).
	
	\item Using previously discussed results, make this operation unitary.
	
	\item Apply the rotation algorithm previously discussed with the unitary we just derived.
	
	\item Decompute the angle used.
\end{enumerate}

\subsection{Angle Estimation}

Let \(\ket{\psi} = \sin(\theta)\ket{\psi'} + \cos(\theta) \ket{0^n}\) for some \(\theta\). Of course to do this, we need to assume that our algorithm is over the reals, and WLOG, we can assume that.

See that for sufficiently small \(\epsilon\), \(-\ln(\epsilon) = O(\frac{1}{\epsilon})\), and \(\frac{1}{\ln(1 - \epsilon)} = O( \frac{1}{\epsilon})\). On \(\frac{-\ln(\frac{1}{2}\epsilon^2)}{2\epsilon^{36}}\) copies of \(\ket{0^n}\), run $U$ and count the fraction of the results that remain \(\ket{0^n}\). This will give us an estimate of \(\cos(\theta)^2\) within \(\epsilon^{18}\) with probability \(1 - \epsilon^2\). This comes from Hoeffding's inequality \cite{Hoe1963} which gives that, for H(n) the calculated \(\cos(\theta)^2\) on n trials:

\[P(|H(n) - \cos(\theta)^2| \geq \epsilon^{18}) \leq 2 e^{-2 \epsilon^{36} n} = 2 e^{-2 \epsilon^{36} \frac{-\ln(\frac{1}{2}\epsilon^2)}{2\epsilon^{36}}} = 2 e^{\ln(\frac{1}{2}\epsilon^2)} = \epsilon^2 \]
	
The calculated \(\epsilon^{18}\) approximation of \(\cos(\theta)^2\) also gives an approximation of \(|\cos(\theta)|\) with error at most \(\epsilon^9\).
	
One last technical detail is that while we can estimate \(|\cos(\theta)|\), we don't know whether \(\cos(\theta)\) is positive or negative. This actually matters because we are using controlled applications of $U$. For instance, if \(\cos(\theta)^2 = 1\), then \(\theta \in \{0, \pi\}\), but we don't know which. Then if we have state \(\ket{+}\ket{0^n}\) and controlled apply $U$, we will get the state \(\ket{-}\ket{0^n}\) if \(\theta = \pi\), but will get state \(\ket{+}\ket{0^n}\) if \(\theta = 0\). So, we need to add one more step to our protocol to figure out whether \(\cos(\theta)\) is positive or negative.
	
The basic idea comes from above. We first prepare a control bit in the state \(\frac{|\cos{\theta}|}{\sqrt{1 + \cos(\theta)^2}}\ket{0} + \frac{1}{\sqrt{1 + \cos(\theta)^2}}\ket{1}\), then controlled on this apply $U$ to \(\ket{0^n}\) and then if the second register is still in the \(\ket{0^n}\) state, the first bit will either be in the \(\ket{+}\) state if \(\cos(\theta) > 0\), or the \(\ket{-}\) state if \(\cos(\theta) < 0\). Of course, we will have to look at this several times to get a high probability of seeing this result, and there is some error in our measured angle we will have to worry about.
	
If we perform this measurement \(\frac{2}{\epsilon^4}\) times, with high probability we will correctly measure the sign if the angle is far enough from \(\frac{\pi}{2}\) to matter. To be specific, if \(|\cos(\theta)| < \epsilon\), then we actually don't care about the sign because we are already close enough to the angle without guessing the sign. That is, we already have our approximately orthogonal computing $U$ for the next step, and don't need to find it. Otherwise, with probability at most \((1 - \epsilon^2)^n\) we will see the result at least once, and for sufficiently small epsilon 
	
\[\log_{1 - \epsilon^2}(\epsilon) = \frac{\ln(\epsilon^2)}{\ln(1-\epsilon^2)} \leq \frac{1}{\epsilon^4} = n \]
	
Thus
	
\[(1 - \epsilon^2)^n \leq (1 - \epsilon^2)^{\log_{1 - \epsilon^2}(\epsilon)} = \epsilon\]
	
Thus with probability at least \(1 - \epsilon\) we will measure the sign of \(\cos(\theta)\).

Now, this measurement can be wrong since we might not have prepared the state exactly right since we don't have \(\cos(\theta)^2\) exactly. However, since we are within \(O(\epsilon^9)\) of \(|\cos(\theta)|\), and we are only considering angles further then \(\epsilon\) from \(\frac{\pi}{2}\), there is an additional \(O(\epsilon^4)\) probability that we make the measurement wrong.

This is because error is increased by up to a \(\frac{1}{\epsilon^2}\) factor from the probability that we measure the \(\ket{0^n}\) state in the first place being as small as \(\epsilon^2\). Then the probability of error is proportional to the square of the sign of the error in the angle. An \(O(\epsilon^9)\) approximation of \(\cos(\theta)\) gives an \(O(\epsilon^{4.5})\) approximation of \(\theta\). Then the sine squared of this error is \(O(\epsilon^9)\). So the difference of the state produced is at most \(O(\epsilon^{4.5})\). But this is small relative to \(O(\epsilon)\).
	
\subsection{Approximate Unitary $V$ for Calculating \(\ket{\psi'}\)}

To construct this $V$, it suffices to give a way to calculate \(\ket{\psi'}\) approximately from \(\ket{0^n}\) without producing any garbage by an earlier result. It suffices to only calculate \(\ket{\psi'}\) approximately because this computation will only be called a constant number of times by a subroutine. Each time will only increase the trace distance from what should have been calculated by at most \(O(\epsilon)\), so the resulting $V$ will still give the correct result with error \(O(\epsilon)\).

Now I give an algorithm that takes \(\ket{0^n}\) to within \(\epsilon\) trace distance of \(\ket{\psi'}\) without producing any garbage, if \(\sin(\theta) < \epsilon\). If \(\sin(\theta) < \epsilon\), as measured earlier, then we don't do anything and we are already within \(\epsilon\) trace distance of the expected state.

Otherwise, prepare \(\frac{1}{\epsilon^2}\) auxiliary bits in the 0 state. Then apply $U$, and conditioned on the result still being \(\ket{0^n}\), flip the first control bit. Then conditioned on it being 1, apply $U$ again. Similarly, conditioned on the result still being in this state, we flip the second bit and apply $U$ again, and so on. After \(\frac{\ln(\epsilon)}{\ln(1 - \epsilon)} < \frac{1}{\epsilon^2}\) iterations, the amplitude of the \(\ket{10^{n-1}}\) is at most \(\epsilon\). Now, we just need to rotate the working space bits back.
	
Well, we know that the amplitude where the first bit is 0 is just \(\sin(\theta)\), and the amplitude that the first bit is 1, given the second bit is 0, is \(\cos(\theta)\sin(\theta)\). Well, we know \(\theta\) (approximately), so we can just rotate by the appropriate angle to make the first bit 0 given the second bit is 0. Then controlled on the second bit being 1, we flip the first one. Then we know the amplitude that the second bit is 0, and the second bit is 1 given the third bit is 0, and we can apply that rotation. And so on.

$$
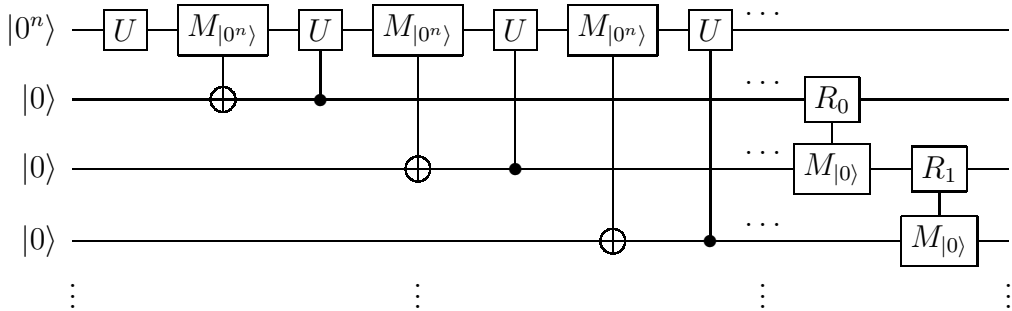
\Qcircuit @C=1em @R=.7em {
	\lstick{\ket{0^n}} & \gate{U} & \gate{M_{\ket{0^n}}} \qwx[1] & \gate{U} & \gate{M_{\ket{0^n}}} \qwx[2] & \gate{U} & \gate{M_{\ket{0^n}}} \qwx[3] & \gate{U} & \ustick{\ldots} \qw & \qw & \qw & \qw \\
	\lstick{\ket{0}} & \qw & \targ & \ctrl{-1} & \qw & \qw & \qw & \qw & \ustick{\ldots} \qw & \gate{R_0} & \qw & \qw \\
	\lstick{\ket{0}} & \qw & \qw & \qw & \targ & \ctrl{-2} & \qw & \qw & \ustick{\ldots} \qw & \gate{M_{\ket{0}}} \qwx[-1] & \gate{R_1} & \qw \\
	\lstick{\ket{0}} & \qw & \qw & \qw & \qw & \qw & \targ & \ctrl{-3} & \ustick{\ldots} \qw & \qw & \gate{M_{\ket{0}}} \qwx[-1]& \qw \\
	\vdots & & & & \vdots & & & & \vdots &  &  & \vdots \\
}$$

Now, our angle isn't exactly, \(\cos(\theta)\) is off by up to an \(\epsilon^9\) amount. Now I need to bound how much error each of these rotations has. Well, the angle to rotate by is dependent on \(\cos(\theta)^n\sin(\theta)\), and \(\sin(\theta)\sum_{i=0}^{n-1}\cos(\theta)^i = \sin(\theta) \frac{1 - \cos(\theta)^n}{1 - \cos(\theta)} \), the amplitude of the two states we are trying to rotate. Then the angle we want to rotate by is
	
\[\theta'_n = \cos^{-1}\left(\frac{\cos(\theta)^n\sin(\theta)}{\sin(\theta)\sqrt{(\frac{1 - \cos(\theta)^n}{1 - \cos(\theta)})^2 + \cos(\theta)^{2n} } } \right) = \cos^{-1}\left(\frac{\cos(\theta)^n}{\sqrt{(\frac{1 - \cos(\theta)^n}{1 - \cos(\theta)})^2 + \cos(\theta)^{2n} } } \right) \]
	
Following Taylor's theorem, we can bound the error of \(\theta_n'\) by its derivative with respect to \(x = \cos(\theta)\). The derivative is
	
\[\frac{-\sqrt{(\frac{1 -x^n}{1 -x})^2 +x^{2n} } }{\frac{1 -x^n}{1 -x}} \left(\frac{1 -x^n}{1 -x} nx^{n-1} +x^n\left(\frac{-nx^{n-1}(1 -x) + 1 - x^n }{(1 -x)^2} \right) \right) \]

Now to bound this, we realize that since \(\sin(\theta) > \epsilon\), \(\cos(\theta)^2 < 1 - \epsilon^2\), thus \(\cos(\theta) < 1 - \epsilon^3\). Then we can bound the absolute value of the above by

Which absolute value is at most
	
\[2(\frac{n}{\epsilon^3} + \frac{n}{\epsilon^3} + 2 \frac{1}{\epsilon^6}) \leq \frac{16}{\epsilon^6} = O(\frac{1}{\epsilon^6})  \]
	
So the total error is going to be less then this derivative times the max error, \(\epsilon^9\). Thus the error in the angle is at most \(O(\epsilon^3)\). the trace distance from applying this angle and the correct angle then is only the sine of this angle, which is strictly less than this angle, so after each of these rotations, we only introduce an \(O(\epsilon^3)\) error in trace distance from the desired rotation which perfectly cleans up the state.

We apply at most \(\frac{1}{\epsilon^2}\) rotations, so the total trace distance from the state that fixes the garbage bits is \(O(\epsilon)\). And the amplitude on the final state that can't be cleaned with one of these rotations is at most \(\epsilon\). So the prepared state is \(O(\epsilon)\) far in trace distance from the expected prepared orthogonal state.

One interesting thing to note in this algorithm is that in preparing this state, we only ever apply \(U\) to \(\ket{0^n}\). So in fact, we don't even need $U$ to be unitary, it just needs to reset its auxiliary bits on input \(\ket{10^{n-1}}\).

\subsection{Applying the Rotation Protocol to V}

Apply protocol for sending to \(\ket{0^n}\) to \(\ket{\psi}\) and leaves \(\ket{0^n}\) unchanged. To do this, we use the above algorithm for getting the unitary that takes \(\ket{0^n}\) to \(\ket{\psi'}\) with trace distance error \(O(\epsilon)\). This new unitary is only called a constant number of times, so the result still is off from the correct answer with trace distance \(O(\epsilon)\).
	
\subsection{Cleaning Up Working Bits Used to Calculate \(\theta\)}

Finally, we clean up the first step's calculation of \(\theta\). This can just be done by running the computation in reverse. This works because for the majority of calculated \(\theta\), the result is within \(\epsilon\) trace distance of the desired state, and only an \(\epsilon^2\) fraction of the calculated \(\theta\) is farther than that. 

To make this explicit, for potential calculation of \(\theta\), \(\theta'\), let \(S_{\theta'}\) be the set of states that during calculation would have yielded \(\theta'\), and for \(x \in S_{\theta'}\) let \(\alpha_x\) be the amplitude on that state. Then the last two steps are a function \(f_{\theta'}: \R^{n+m} \rightarrow \R^{n+m}\), and the result of the algorithm at this point is:

\[\sum_{\theta'} (\sum_{x \in S_{\theta'}} \alpha_x \ket{x}) \ket{\theta'} \ket{f_{\theta'}(y)}\]

for input y. We also know that for some unitary $W$ which swaps \(\ket{0^n}\) and \(\ket{\omega}\), does nothing for any orthogonal input, and leaves all auxiliary bits at 0, for \(\theta' \in (\theta - \epsilon^9, \theta + \epsilon^9)\),  \(\ket{f_{\theta'}(y)}\) is within trace distance \(O(\epsilon)\) of \(W\ket{y}\). And finally, if we let \(R\) be the set of calculated \(\theta'\) within \(\epsilon^9\) of \(\theta\), or loosely \(R = (\theta - \epsilon, \theta + \epsilon)\). Then we started by showing that

\[\sum_{\theta' \in R} \sum_{x \in S_{\theta'}} \alpha_x^2 > 1 - \epsilon^2\]
Or
\[\sum_{\theta' \in R^c} \sum_{x \in S_{\theta'}} \alpha_x^2 < \epsilon^2\]

So first, we see that if we consider the \(\epsilon\) fraction of the inputs that are far away from the real \(\theta\), modifying them to have their result be the unitary we are approximating is only \(O(\epsilon)\) trace distance away. This is because the trace distance of pure states \(\ket{a}, \ket{b}\) is just \(\sqrt{1 - \bra{a}\ket{b}}\), and the inner product of this perturbed state with the original is just

\[
\begin{split}
&(\sum_{\theta' \in R} (\sum_{x \in S_{\theta'}} \alpha_x \bra{x}) \bra{\theta'} \bra{f_{\theta'}(y)} + \sum_{\theta' \in R^c} (\sum_{x \in S_{\theta'}} \alpha_x \bra{x}) \bra{\theta'} \bra{f_{\theta'}(y)} )\\
& (\sum_{\theta' \in R} (\sum_{x \in S_{\theta'}} \alpha_x \ket{x}) \ket{\theta'} \ket{f_{\theta'}(y)} + \sum_{\theta' \in R^c} (\sum_{x \in S_{\theta'}} \alpha_x \ket{x}) \ket{\theta'} W \ket{y} ) \\
= & \sum_{\theta' \in R} (\sum_{x \in S_{\theta'}} \alpha_x^2) + \sum_{\theta' \in R^c} (\sum_{x \in S_{\theta'}} \alpha_x^2 ) \bra{f_{\theta'}(y)} W\ket{y}   \\
= & 1 - \sum_{\theta' \in R^c} (\sum_{x \in S_{\theta'}} \alpha_x^2) + \sum_{\theta' \in R^c} (\sum_{x \in S_{\theta'}} \alpha_x^2 ) \bra{f_{\theta'}(y)} W\ket{y}  \\
> & 1 - 2\sum_{\theta' \in R^c} (\sum_{x \in S_{\theta'}} \alpha_x^2) \\
> & 1 - 2\epsilon^2 \\
\end{split}
\]

So the trace distance of these states is less than
\[\sqrt{1 - (1 - 2\epsilon^2)^2} = \sqrt{4\epsilon^2 - 4\epsilon^4} = 2\epsilon\sqrt{1 - \epsilon^2} = O(\epsilon)\]

Now we perturb it again, this time only modifying the inputs that are close to \(\theta\). There are a lot more of these, but each of them is only trace distance \(O(\epsilon)\) from \(W\ket{y}\). Intuitively, this should imply that the combination of these with an unperturbed state should also have that trace distance. But to make this concrete, lets calculate the trace distance explicitly. To say \(\ket{f_{\theta'}(y)} \) and \(W\ket{y}\) have trace distance \(O(\epsilon)\) means that for some constant $K$: \(\bra{f_{\theta'}(y)}W\ket{y} > \sqrt{1 - K\epsilon^2}\). Thus when we look at the inner product of the last state with the perturbed state, we get:

\[
\begin{split}
&(\sum_{\theta' \in R} (\sum_{x \in S_{\theta'}} \alpha_x \bra{x}) \bra{\theta'} \bra{f_{\theta'}(y)} + \sum_{\theta' \in R^c} (\sum_{x \in S_{\theta'}} \alpha_x \bra{x}) \bra{\theta'} \bra{y} W^{-1} )\\
& (\sum_{\theta' \in R} (\sum_{x \in S_{\theta'}} \alpha_x \ket{x}) \ket{\theta'} W \ket{y} + \sum_{\theta' \in R^c} (\sum_{x \in S_{\theta'}} \alpha_x \ket{x}) \ket{\theta'} W \ket{ y} ) \\
= & \sum_{\theta' \in R} (\sum_{x \in S_{\theta'}} \alpha_x^2) \bra{f_{\theta'}(y)} W \ket{y} + \sum_{\theta' \in R^c} (\sum_{x \in S_{\theta'}} \alpha_x^2 ) \\
> & \sum_{\theta' \in R} (\sum_{x \in S_{\theta'}} \alpha_x^2) \sqrt{1 - K\epsilon^2} + \sum_{\theta' \in R^c} (\sum_{x \in S_{\theta'}} \alpha_x^2 ) \\
> & \sqrt{1 - K\epsilon^2} \\
\end{split}
\]

So the trace distance of these states is less than
\[\sqrt{1 -  \sqrt{1 - K\epsilon^2}^2} = \sqrt{K}\epsilon \]

Therefore, the trace distance fixing after fixing these states is \(O(\epsilon)\). Therefore, the trace distance from the original state, to what we wanted to calculate is \(O(\epsilon)\). That is,

\[tr(\sum_{\theta'} (\sum_{x \in S_{\theta'}} \alpha_x \ket{x}) \ket{\theta'} \ket{f_{\theta'}(y)}, \sum_{\theta'} (\sum_{x \in S_{\theta'}} \alpha_x \ket{x}) \ket{\theta'} W\ket{y}) = O(\epsilon) \]

And what we wanted to calculate is independent of the first two registers, so when we undo the counting procedure on it, we perfectly clean up the input bits and have calculated what we want. Thus when we apply this to our actual state which is \(O(\epsilon)\) close to this state, we end up \(O(\epsilon)\) close to cleaning up the input bits and calculating what we want.

\section{Future Directions}

There are a few open questions about this problem.

Even though this algorithm is polynomial in \(\frac{1}{\epsilon}\), the polynomial is huge, \(O(\frac{-\ln(\epsilon)}{\epsilon^{36}})\) since angle estimation is by far the slowest part. Its very rare for simple algorithms to require such large run times. Some of this could be potentially improved with a more careful analysis. But likely, to get much better, we need a more efficient technique to calculate \(\ket{\omega'}\). Can this be done?

Although this result only approximates the correct answer, we haven't ruled out an exact computation that runs with constant queries. This can be done easily in a few special cases discussed above, but seems very difficult in general. Is It?

This algorithm only works for transforming 2 orthogonal input states to 2 orthogonal output states. The algorithm seemed to rely on the fact that it was only 2. Can we extend it to work on a bigger set of input and output states?

\printbibliography

\end{document}